\documentstyle[epsfig,osa,aps]{revtex}

\begin{document}

\title{Lagrangian Velocity Correlations 
and Absolute Dispersion in the Midlatitude Troposphere}
\author{Jai Sukhatme}
\address{Advanced Study Program, National Center for Atmospheric Research, Boulder, CO}
\date{\today}
\maketitle


\begin{abstract}
Employing daily wind data from the ECMWF, we perform passive particle advection
to estimate the Lagrangian velocity correlation functions (LVCF) associated with the midlatitude 
tropospheric flow. In
particular we decompose the velocity field into time mean and transient (or eddy) components to
better understand the nature of the LVCF's. A closely related quantity, 
the absolute dispersion (AD) is also examined. \\

Given the anisotropy of the flow, meridional and zonal 
characteristics are considered separately. 
The zonal LVCF is seen to be non-exponential. In fact, 
for intermediate timescales it can either be interpreted as a power law of the 
form $\tau^{-\alpha}$ with $ 0<\alpha<1$ or as the sum of exponentials with differing timescales -
both interpretations being equivalent. More importantly the 
long time correlations in the zonal flow result in
a superdiffusive zonal AD regime. On the other hand, the meridional LVCF decays rapidly to zero. 
Before approaching zero the meridional LVCF shows
a region of negative correlation - a consequence of the presence of planetary scale Rossby 
waves. 
As a result the meridional AD, apart from showing 
the classical asymptotic ballistic and diffusive regimes, displays transient 
subdiffusive behaviour. \\
\end{abstract}


\section{Introduction}
Being a quantity of fundamental interest in the statistical characterization of 
a turbulent flow, the Lagrangian velocity correlation function (LVCF)
has been the subject of detailed study \cite{Hinze}, \cite{Monin}. Indeed, if we 
view the midlatitude free troposphere to be in a non-ideal turbulent state \cite{Shep_1},
\cite{Shep_2}, it is of natural interest inquire into the nature its LVCF. 
Similarly, apart from being a measure of transport and its close relation to
the LVCF \cite{Taylor},\cite{Monin}, the absolute dispersion (AD) is known to yield information
about the structure of the underlying flow \cite{BG},\cite{Zas}. \\

Ofcourse there exist other measures,
such as the relative dispersion (or Lyapunov exponents for
smooth flows) and finite scale analyses (or finite size Lyapunov exponents), which are
a part of the
complete statistical characterization of a flow. In particular, finite scale statistics are
argued to be useful when the available range of scales is small and there is the possibility
of crossover effects from differing regimes \cite{Boffetta}
\footnote{I thank Referee A for the discussion of finite scale statistics.\\}.
Our approach is slightly different from earlier modelling studies of large scale
atmospheric dispersion which employ a Langevin equation, equivalently assuming an exponential LVCF,
to parameterize the dispersion process \cite{Giff1},\cite{Giff2} (a discussion of
such an approach can be found in \cite{Sawford}).
In the present note we will only
consider the LVCF and AD, focussing
their characterization and complementary nature. Also, we will see the
accordance of these statistics with simpler physically relevant systems. \\

For computational purposes, daily pressure level global T63 (192 $\times$ 96, 17 levels) resolution 
wind data 
from the ECMWF is used as the 3D 
advecting velocity field. 
We computed the trajectories (and record the velocity along these trajectories) of an 
ensemble of passive particles. The trajectories were
computed in latitude, longitude and pressure coordinates by a standard fourth order 
Runge-Kutta scheme and the velocity fields 
were interpolated in a linear fashion. Averaging is done with respect to an ensemble 
(denoted by {\sf S}) of passive particles which remain in the free troposphere, i.e. particles
which escape into the boundary layer or into the stratosphere are excluded from the 
statistics \footnote{Experiments were performed with an initial set of 20,000 and 30,000 particles,
no discernable difference in the statistics was noticed in these two situations.\\}. \\


\section{Lagrangian Velocity Correlation Functions}
\subsection{Zonal LVCF}
Denoting the zonal velocity by $u(\lambda,\phi,p,t)$, the zonal LVCF ($R_u(\tau)$) is defined as

\begin{equation}
R_u(\tau) = \frac{< u(\vec{x}(t+\tau))~u(\vec{x}(t)) >_{\sf S} }{< u(\vec{x}(t))^2 >_{\sf S}}
\label{1}
\end{equation}
Here $\vec{x}(t)$ represents the trajectory of an individual member of {\sf S}. 
To account for the inhomogeneity of the flow and to focus our attention the midlatitudes, 
{\sf S} is chosen to comprise of N members located 
randomly such that they initially satisfy $0^\circ < \lambda({\sf S}) < 360^\circ$, 
$35^\circ \le \phi({\sf S}) \le 55^\circ$ and $400 \textrm{mb} \le p({\sf S}) \le 700 \textrm{mb}$,
where $\lambda,\phi,p$ represent the longitude, latitude and pressure coordinates. Note that 
no restriction is placed on the trajectories, i.e. the statistics presented are {\it not conditional}
on the particles remaining in the midlatitudes for all times. \\

\begin{figure}
\begin{center}
\epsfxsize=12.0 cm
\epsfysize=12.0 cm
\leavevmode\epsfbox{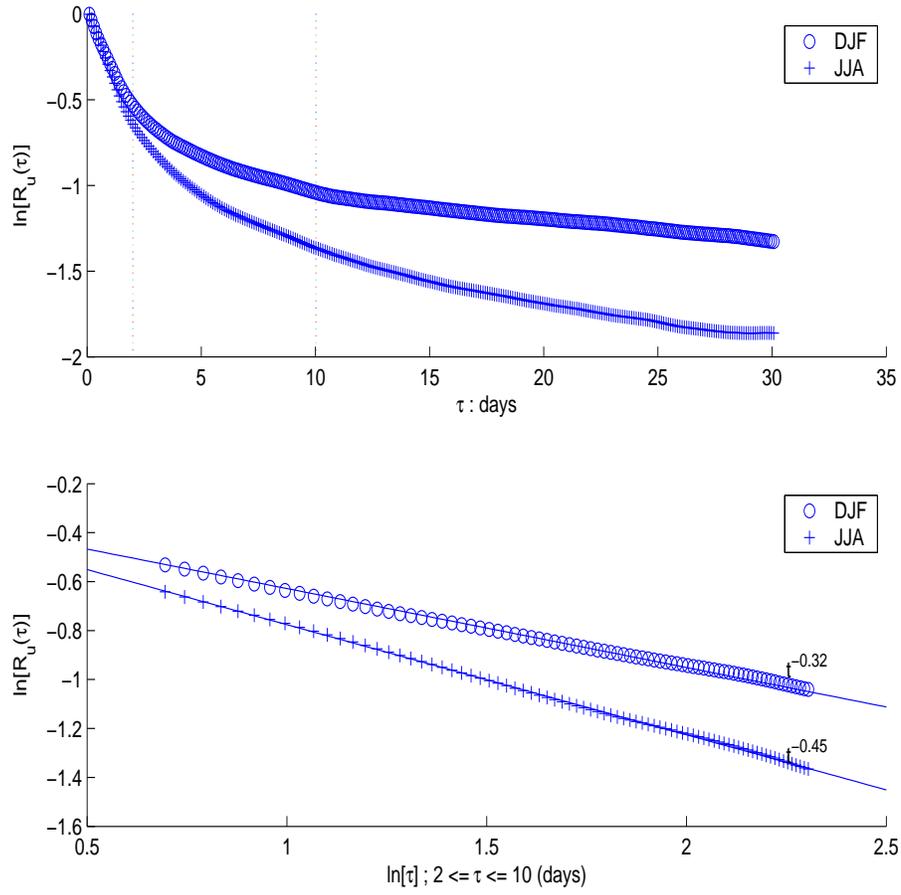}
\end{center}
\caption{Upper Panel : Zonal LCVF. Lower Panel : Possible power law behaviour at intermediate 
timescales. Though a sum of exponential processes would result in similar behaviour at
intermediate scales. }
\label{fig:fig1}
\end{figure}

As can be seen in the upper panel of 
Fig. \ref{fig:fig1}, which shows $\log[R_u(\tau)]$ 
for the winter and summer seasons, the zonal LVCF is clearly non-exponential. 
This is in agreement with studies on 2D \cite{Pasquero} and geostrophic 
\cite{Pecseli} turbulence, but stands in contrast to investigations of
3D turbulence \cite{Mordant},\cite{Pope}. 
Further, for intermediate values of $\tau$ (i.e. $2 < \tau < 10$ days),
it appears that (lower panel of Fig. \ref{fig:fig1})                                                 
$R_u(\tau) \sim \tau^{-\alpha}$
with $0 < \alpha < 1$ ($\alpha = 0.32 ~\textrm{and}~ 0.45$ for DJF and JJA respectively).
The issue of whether the correlation function is indeed a power law is known to be a 
delicate matter. Indeed any monotonically continuous function, such as the 
aforementioned power law, can be expressed (over a given range of scales) as the discrete sum of 
exponentials - see Berglund \cite{Berg} for definitions and details - 
a fact which has been elegantly 
utilized in the analysis of queues in communication networks \cite{Anja},\cite{Star}. 
Whether the zonal LVCF is the sum of two (or more) Ornstein-Uhlenbeck processes with 
differing timescales - as is argued to be an effective parameterization of the LVCF in
2D turbulence \cite{Pasquero} - or, if it indeed shows power-law scaling is
practically undecidable especially if the behaviour is seen over a small range of scales 
\footnote{Comments of Referee B are acknowledged as they provoked a more careful look at 
this issue.\\}.
On the other hand, what is certain is the enhancement of the Lagrangian correlation
time \cite{BG} ($T_u = {\int_0}^{\infty}~ R_u(t) dt$). Indeed, as will be observed,
the persistent correlations in the zonal velocity field result in a superdiffusive
AD regime. \\

\subsection{Meridional LVCF}
\begin{figure}
\begin{center}
\epsfxsize=9.5 cm
\epsfysize=8.5 cm
\leavevmode\epsfbox{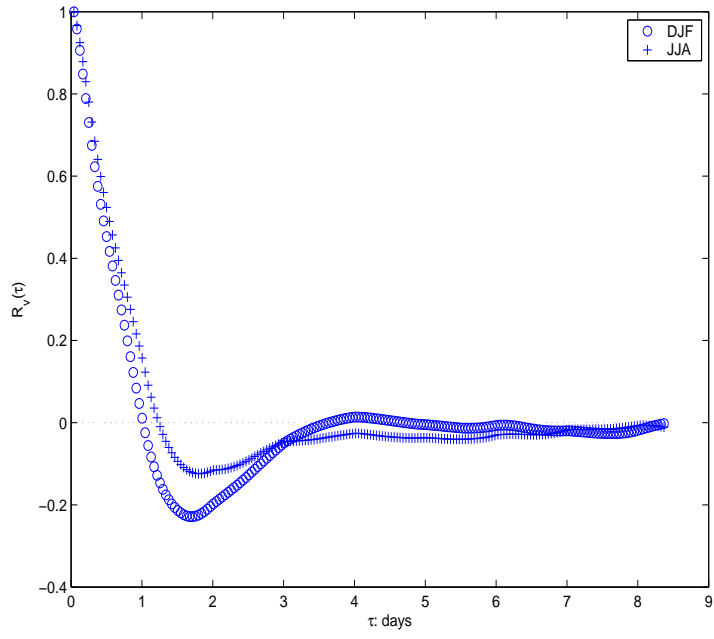}
\end{center}
\caption{Meridional LCVF : Note $R_v(\tau) < 0$ before $R_v(\tau) \rightarrow 0$.}
\label{fig:fig2}
\end{figure}

Defined in a manner similar to Eq. (\ref{1}), but using the meridional component of the
velocity field ($v(\lambda,\phi,p,t)$) we compute the meridional LVCF ($R_v(\tau)$). As
is seen in Fig. \ref{fig:fig2}, $R_v(\tau)$ decays to zero in one week.
Interestingly, we notice a pronounced anticorrelation, i.e. $R_v(\tau) < 0$ before the
final $R_v(\tau) \rightarrow 0$ behaviour. On comparison with 2D \cite{Pasquero} and
geostrophic turbulence \cite{Pecseli} we find this feature to be unique to the present
situation. This is easily explained when one takes into account the latitudinal
restriction imposed by the rotation of the planet \cite{Rhines}. Specifically, 
in the absence
of strong diabatic or frictional effects, the conservation of potential
vorticity gives rise to large scale Rossby waves \cite{Pedlosky}. We argue that the 
meridional oscillatory motion implied by the Rossby waves is responsible for the 
aforementioned anticorrelation. \\

\subsection{Eddy and Time Mean LVCF's}
To gain some insight into the connection
between the nature of the LVCF's and
the structure of the tropospheric flow, we partition the daily data into time mean and transient
components. Specifically, the time mean is $\hat{u}(\vec{x}) = (1/T){\int_0}^T~u(\vec{x},t) dt$
(where $T$ is the
duration of the entire season) and the transient or
eddy component is defined as $u'(\vec{x},t) = u(\vec{x},t) - \hat{u}(\vec{x})$.
Even though the tropospheric
flow does not posses a clear spectral gap, i.e. there is a near continuum of active (temporal
and spatial) scales,
we expect the above decomposition to separate processes which vary on scales that are farthest apart
\cite{Black}. \\

\begin{figure}
\begin{center}
\epsfxsize=11 cm
\epsfysize=12 cm
\leavevmode\epsfbox{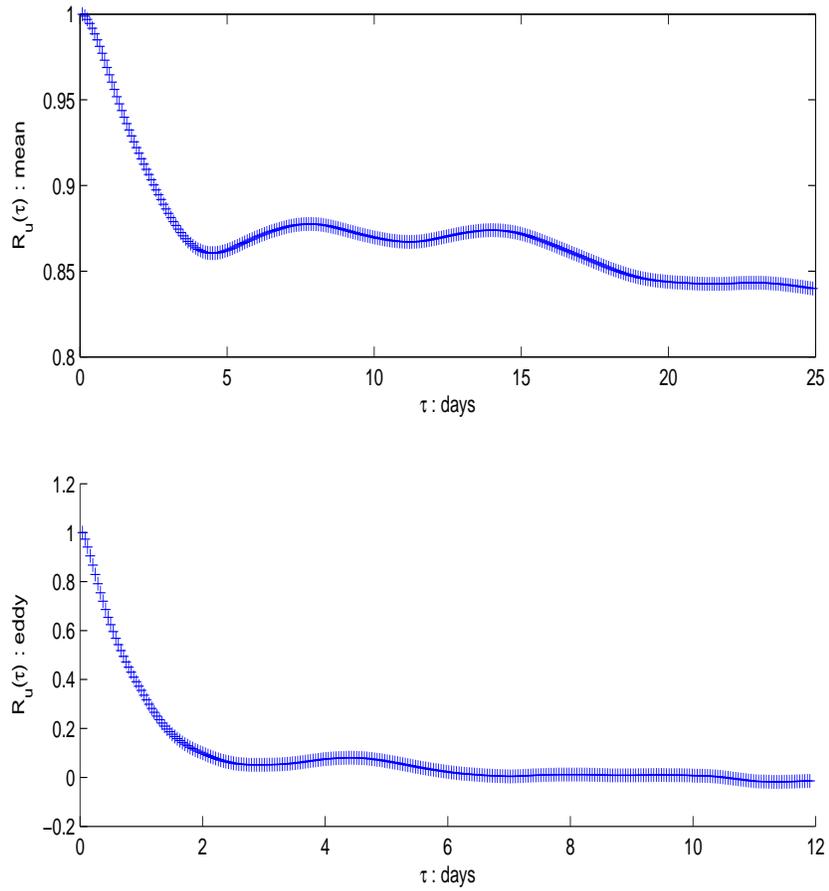}
\end{center}
\caption{Upper Panel : Time Mean Zonal LCVF (DJF data). Lower Panel : Eddy Zonal LCVF. Note the 
different timescales in the two panels. Also, by about one week the eddy correlations have almost
completely died out wheras the mean flow is still strongly correlated. JJA data (not shown) behaves in a 
qualitatively similar manner.}
\label{fig:fig3}
\end{figure}

\begin{figure}
\begin{center}
\epsfxsize=11 cm
\epsfysize=12 cm
\leavevmode\epsfbox{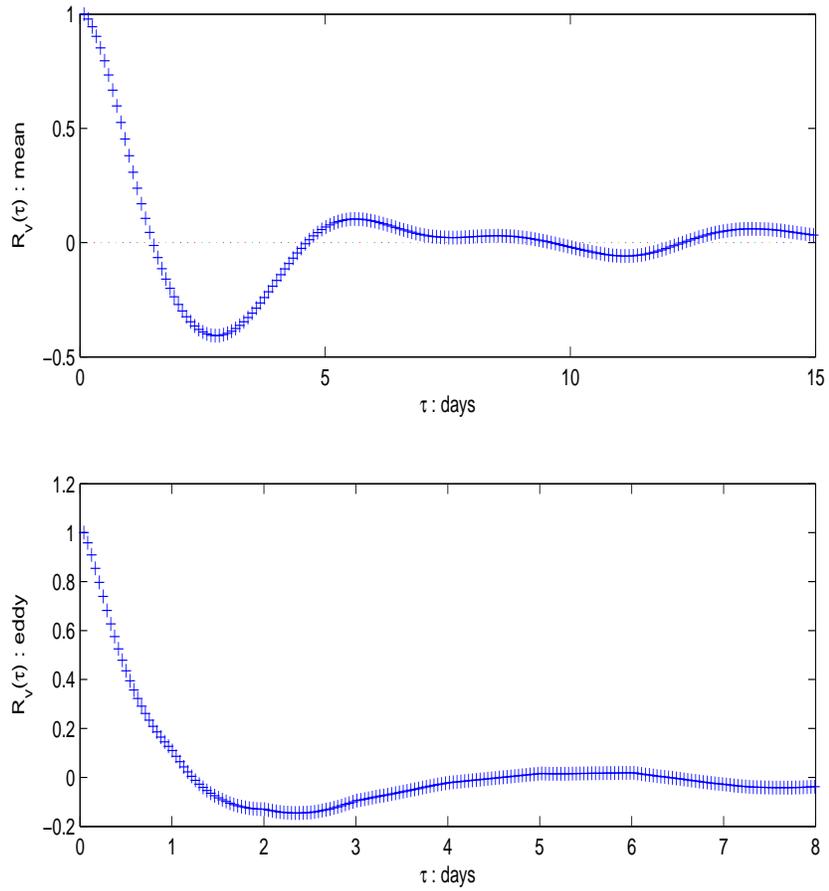}
\end{center}
\caption{Upper Panel : Time Mean Meridional LCVF (DJF data). Lower Panel : Eddy Meridional LCVF. 
Once again, note the 
different timescales in the two panels. JJA data (not shown) behaves in a qualitatively similar manner.}
\label{fig:fig4}
\end{figure}

Fig. \ref{fig:fig3} and Fig. \ref{fig:fig4} show the zonal and meridional LVCF's computed 
by exculsively utilizing the Eulerian time 
mean (upper panels) and Eulerian eddy fields (lower panels). In spite of the crudeness of our partition,
the similarity between the eddy LVCF's in both cases is evident. Indeed, apart from the slight 
anticorrelation retained in the meridional eddy LCVF, both $R_{u'}(\tau)$ and $R_{v'}(\tau)$
are rapidly decaying functions with a timescale of the order of a couple of days. \\

On the other hand the zonal and meridional time mean LVCF's are strikingly different. 
Where $R_{\hat{u}}(\tau)$ is strongly correlated on long timescales - something we would expect 
from a slowly varying zonal jet - 
$R_{\hat{v}}(\tau)$ represents oscillatory motion as induced by a large scale wave. 
Moreover, comparing the behaviour of $R_{v'}(\tau)$ and $R_{\hat{v}}(\tau)$ leads us
primarily attribute the anticorrelation observed in $R_v(\tau)$ to the time mean
component of the
flow. 
Indeed, behaviour consistent with these results has been observed in studies 
of balloon trajectories in the Southern Hemisphere \cite{Morel-D} (see especially their Figs. 9, 10 
and the discussion regarding the timescales involved in the 
definition of stationary and transient components of the flow). \\

\section{Absolute Dispersion}
The absolute dispersion (AD) is defined as,

\begin{equation}
\textrm{AD}(t) = < (x(t) - x(0))^2 >_{\sf S}
\label{2}
\end{equation}
For ideal (i.e. isotropic, homogenous, stationary and zero time mean) flows, we have \cite{Taylor},
\cite{Monin},

\begin{equation}
\textrm{AD}(t) = 4E {\int_0}^t R(\tau)~(t-\tau) ~d\tau ~;~ E = \textrm{kinetic energy}
\label{3}
\end{equation}
The short time limit (i.e. $\tau \rightarrow 0$) of the above yields ballistic motion
whereas when $T_u$ is finite, the long time limit (i.e. $\tau >> T_u$) yields diffusive
behaviour \cite{Hinze}. The presence of other exponents, i.e. AD(t) $\sim t^{\gamma} ~;~ \gamma \neq 1$,
is referred to as anomalous diffusion \cite{BG}. Before displaying the results, 
let us fix 
some notation. We denote the total AD by $A(t)$, i.e.
$A(t)=< (x(t) - x(0))^2 + (y(t) - y(0))^2 + (z(t) - z(0))^2 >_{\sf S}$, where $x,y,z$ are
cartesian coordinates. Further the individual components of the AD are denoted by
$A_i(t)$ where $i$ represents a coordinate, for eg. $A_x(t) = < (x(t) - x(0))^2 >_{\sf S}$. \\

\begin{figure}
\begin{center}
\epsfxsize=9.5 cm
\epsfysize=8.5 cm
\leavevmode\epsfbox{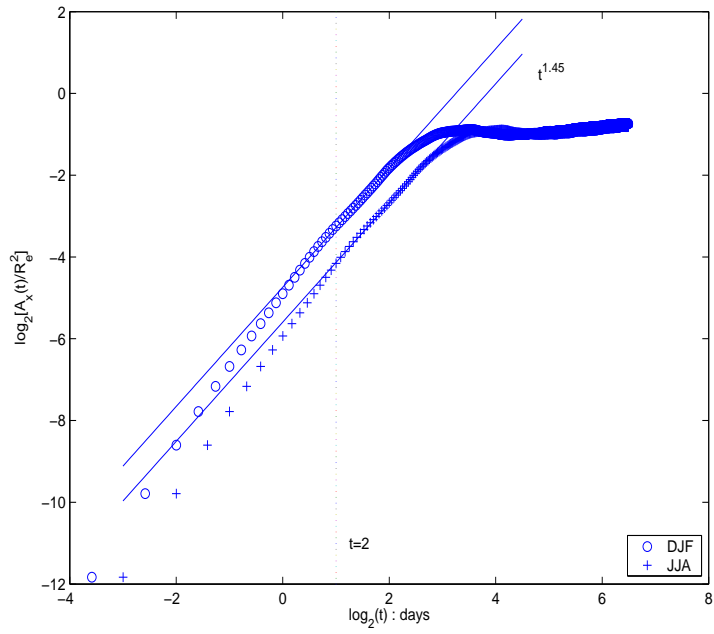}
\end{center}
\caption{Zonal AD. Ballistic $\rightarrow$ Superdiffusive $\rightarrow$ Saturation}
\label{fig:fig5}
\end{figure}

\begin{figure}
\begin{center}
\epsfxsize=11 cm
\epsfysize=12 cm
\leavevmode\epsfbox{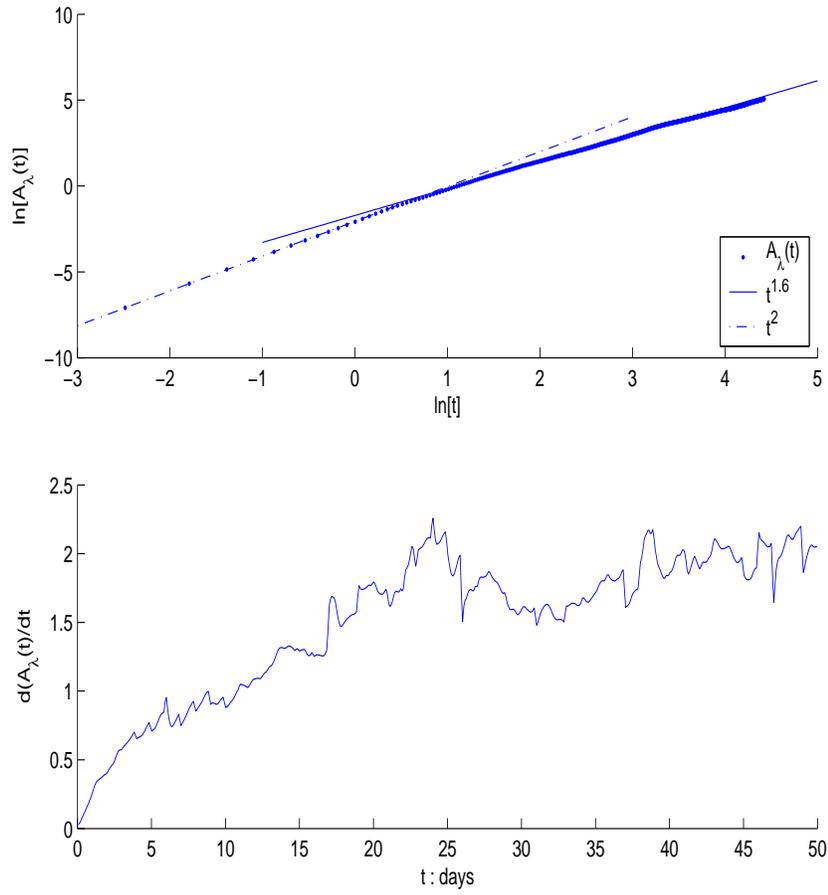}
\end{center}
\caption{Upper Panel : DJF Longitudinal AD (Ballistic $\rightarrow$ Superdiffusive $\rightarrow$ Diffusive).
Lower Panel : $dA_{\lambda}(t)/dt$ Vs. t. Note $dA_{\lambda}(t)/dt \rightarrow$ Const. $\Rightarrow$ 
normal diffusion. }
\label{fig:fig6}
\end{figure}

From Eq. (\ref{3}) we have $R(\tau) \sim \tau^{-\alpha} \Rightarrow \textrm{AD}(t) \sim t^{2-\alpha}$.
Even though the non-ideal nature of the present flow, especially its rich time mean
structure \cite{Shep_1},\cite{Shep_2},
is likely to invalidate the direct applicability of Eq. (\ref{3}) - nonetheless,
from the form of $R_u(\tau)$ (whether one takes it to be a power law or a sum of exponentials, both 
being equivalent from the discussion in the previous section) it is reasonable to expect the 
zonal AD to exhibit 
anomalous behaviour at
intermediate timescales. Indeed as can be seen in Fig. \ref{fig:fig5} (which shows $A_x(t)$ as 
computed using the daily wind data) 
\footnote{$A_y(t)$ is virtually identical to $A_x(t)$. Also, $A_z(t) ~
(\textrm{shown later}~)<< A_x(t)$
hence
$A(t)$ also behaves in the same fashion as $A_x(t)$. \\}, 

\begin{equation}
A_x(t) \sim t^2 \quad 0 < t \le 2 \quad \textrm{[Ballistic]}
\label{4}
\end{equation}
\begin{equation}
A_x(t) \sim t^{\gamma}~;~\gamma=1.45 \quad 2 < t \le 8 \quad \textrm{[Anomalous : Superdiffusive]}
\label{5}
\end{equation}
\begin{equation}
A_x(t) \sim R_e^2~;~R_e=\textrm{Earth Radius}\quad t > 8 \quad \textrm{[Saturation]}
\label{6}
\end{equation}
In order to avoid the effective boundedness of the
domain
we unwrap the longitude and present $A_{\lambda}(t)$ in Fig. \ref{fig:fig6}.
Now the anomalous regime ($t^{\delta} ~;~ \delta = 1.6$) 
lasts from $T_1 < t < T_2$ days ($T_1 \sim 2$ and $T_2 \sim 25$), after which we see the
beginning of
an asymptotic diffusive regime (lower panel of Fig. \ref{fig:fig4}) \footnote{Note that 
we should expect $\delta \neq \gamma$
as $A_{\lambda}(t)$
only involves changes in $\lambda$ whereas $A_x(t)$ is sensetive to both $\lambda$ and $\phi$. \\}.
It is worth mentioning that the anomalous behaviour of $A_{\lambda}(t)$
is in close agreement with laboratory experiments on quasi-geostrophic flows
\cite{Weeks}. Further, zonal superdiffusion has been identified in studies involving large 
amplitude Rossby waves \cite{Flierl} and in more general PV conserving flows (where 
ofcourse, the superdiffusion is along PV contours) \cite{LaCasce} 
\footnote{I thank Referee C for providing references on particle motion
in PV conserving systems.\\}.\\

\begin{figure}
\begin{center}
\epsfxsize=11 cm
\epsfysize=12 cm
\leavevmode\epsfbox{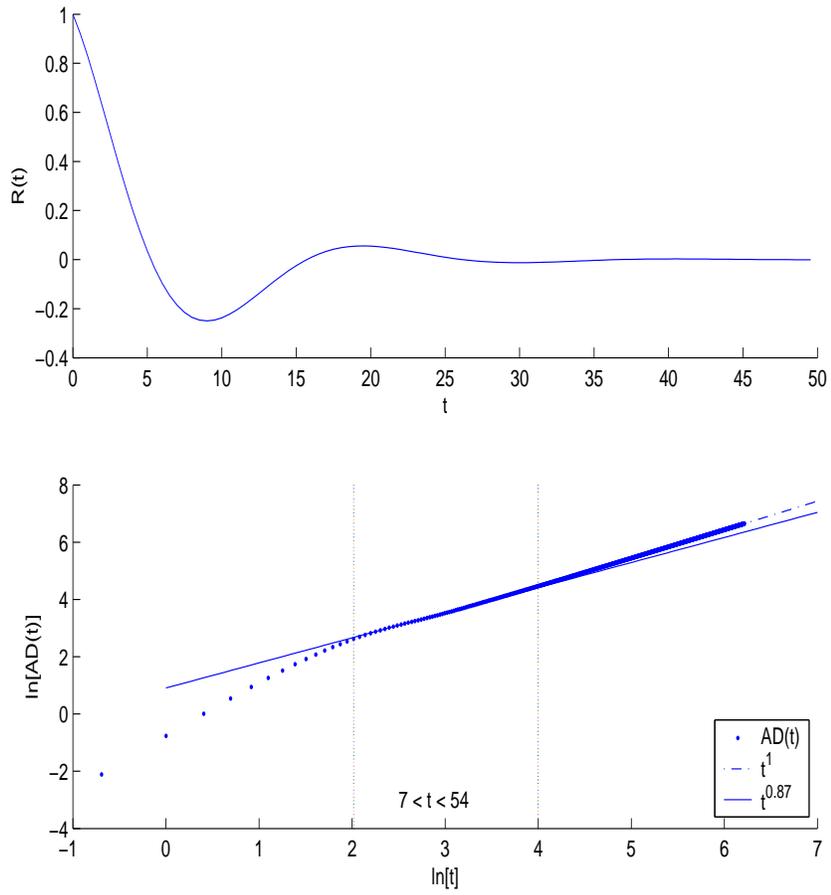}
\end{center}
\caption{Upper Panel : Synthetic LVCF $R(\tau) \sim \textrm{e}^{-\tau/C_1} ~ \cos(\omega \tau)$
($C_1=7 , \omega=0.3 $).
Lower Panel : Induced AD.}
\label{fig:fig7}
\end{figure}

\begin{figure}
\begin{center}
\epsfxsize=12 cm
\epsfysize=12 cm
\leavevmode\epsfbox{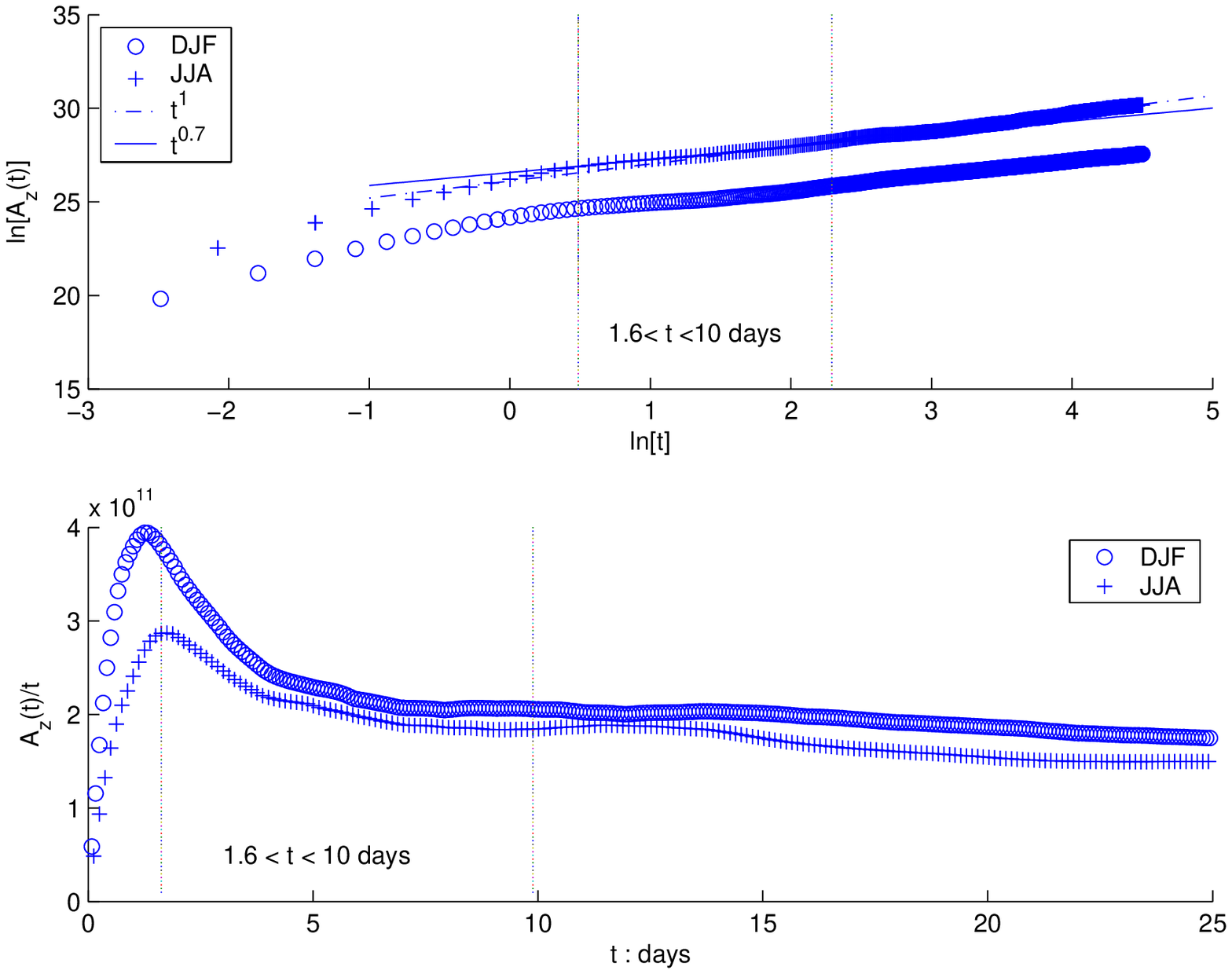}
\end{center}
\caption{Upper Panel : Meridional AD (JJA and DJF curves have been shifted for clarity). 
Ballistic $\rightarrow$ Subdiffusive $\rightarrow$ Diffusive. 
Lower Panel : $A_z(t)/t \sim t \rightarrow A_z(t)/t \sim t^{\beta} ~-1<\beta<0 \rightarrow 
A_z(t)/t \sim$ Const.}
\label{fig:fig8}
\end{figure}

Regarding the meridional AD, apart from the initial ballistic behaviour we expect to see
normal diffusion at large $t$, as $R_v \rightarrow 0$ quite rapidly. 
Once again, we use Eq. (\ref{3}) to get a feel for the meridional AD
at intermediate timescales. Qualitatively approximating $R_v(\tau) \sim \textrm{e}^{-\tau/C_1} ~ 
\cos(\omega \tau)$ (Fig. \ref{fig:fig2}), numerical integration of Eq. (\ref{3}) yields the AD
shown in Fig. \ref{fig:fig7}. Apart from the two asymptotic 
regimes 
we notice transient subdiffusive scaling. This is in accord with results utilizing 
random shear flows, where anticorrelation in the LCVF was associated with 
subdiffusion and even complete trapping in extreme cases \cite{Elliott}. 
Indeed the actual meridional AD, $A_z(t)$ shown in 
Fig. \ref{fig:fig8}, behaves in precisely the same manner, 

\begin{equation}
A_z(t) \sim t^{2} \quad 0 < t \le 2 \quad \textrm{[Ballistic]}
\label{7}
\end{equation}
\begin{equation}
A_z(t) \sim t^{0.7} \quad 2 < t \le 7 \quad \textrm{[Anomalous : Subdiffusive]}
\label{8}
\end{equation}
\begin{equation}
A_z(t) \sim t \quad t > 7 \quad \textrm{[Diffusive]}
\label{9}
\end{equation}

Comparing this behaviour with the meridional AD in strictly PV conserving flows \cite{Flierl},
\cite{LaCasce} we see that in those cases the meridional AD is bounded whereas the violation
of PV conservation at large times (i.e. greater than a week) leads to unbounded normal 
diffusion in the present situation.
Such superdiffusive zonal and subdiffusive meridional 
behaviour has also been recently observed 
in a study of anisotropic drift-wave
turbulence \cite{Basu} - particularly remarkable is the similarity of the anomalous exponents in 
the two situations. \\

\section{Summary}

Employing daily wind data from the ECMWF, we have estimated the zonal and meridional LVCF's
of the midlatitude tropospheric flow. The zonal LVCF is seen to be non-exponential in character.
Physically, given that the midlatitude tropospheric flow has a rich time mean structure 
along with an energetic 
eddy field \cite{Shep_1},\cite{Shep_2} - this observation is not entirely unexpected. 
Moreover, from this perspective
our examination of $R_{\hat{u}}(\tau)$ and $R_{u'}(\tau)$ serves to 
clarify the roles of the time mean and eddy fields respectively. Specifically, the eddy 
field by itself generates an almost exponential rapidly decaying LCVF whereas the time mean 
component - roughly 
a slowly varying unidirectional jet flow \cite{Black} - is seen to be strongly correlated. \\

Apart from decaying to zero in a relatively short time ($\approx 1$ week), the meridional LVCF
exhibits an anticorrelation - i.e. $R_v(\tau) < 0$ before $R_v(\tau) 
\rightarrow 0$. We attribute this anticorrelation to 
the presence of large scale planetary waves - a basic consequence of PV conservation on a rotating planet.
Examining $R_{v'}(\tau)$ we see that the meridional eddy 
LCVF is very similar to its zonal counterpart. Whereas $R_{\hat{v}}(\tau)$ - a manifestation
of the large scale stationary waves - has an oscillatory character and indicates the 
time mean component to be primarily responsible for
the above mentioned anticorrelation in $R_v(\tau)$. \\

As regards the AD 
the point that stands out is the simultaneous existence of superdiffusive and subdiffusive 
anomalous scaling
in the zonal and meridional directions respectively. It must be stressed that we lack a quantitative
relationship between the LCVF and the AD in this non-ideal situation. Nonetheless, a certain 
qualitative basis is provided by super- and sub-diffusive behaviour in ideal turbulent fields
with enhanced (power laws or sums of exponentials depending on ones interpretation) and 
anticorrelated LVCF's respectively
\footnote{Note that this anomalous behaviour is transient i.e., it is flanked
on either side by an asymptotic regime. Even though in the present situation this behaviour 
is supported by the 
nature of the LVCF, it is worth keeping in mind that crossover effects (especially in the 
superdiffusive case) could play a role
in determining the quantitative nature of the anomalous exponents \cite{Boffetta}.\\}.
Finally, given that similar 
behaviour has been observed in drift wave turbulence \cite{Basu}, we are led to 
speculate on the possible universality of this phenomenon in fields where 
(slow) jets and waves co-exist with (fast) eddies. \\

\acknowledgments
Comments by Dr. R. Saravanan are gratefully acknowledged. Also, comments by 
all three referees led to a significant 
improvement in the material presented. This work was
carried out at the National Center for Atmospheric Research which is sponsored by 
the National Science Foundation.

\clearpage

\section{Figure Captions}

\begin{itemize}
\item Figure 1 : Upper Panel : Zonal LCVF. Lower Panel : Possible power law behaviour at intermediate
timescales. Though a sum of exponential processes would result in similar behaviour at
intermediate scales.

\item Figure 2 : Meridional LCVF : Note $R_v(\tau) < 0$ before $R_v(\tau) \rightarrow 0$.

\item Figure 3 : Upper Panel : Time Mean Zonal LCVF (DJF data). Lower Panel : Eddy Zonal LCVF. Note the
different timescales in the two panels. Also, by about one week the eddy correlations have almost
completely died out wheras the mean flow is still strongly correlated. JJA data (not shown) behaves in
a qualitatively similar manner.

\item Figure 4 : Upper Panel : Time Mean Meridional LCVF (DJF data). Lower Panel : Eddy Meridional LCVF. Once again, note the 
different timescales in the two panels. JJA data (not shown) behaves in a qualitatively similar manner.

\item Figure 5 : Zonal AD. Ballistic $\rightarrow$ Superdiffusive $\rightarrow$ Saturation.

\item Figure 6 : Upper Panel : DJF Longitudinal AD (Ballistic $\rightarrow$ Superdiffusive $\rightarrow$ Diffusive).
Lower Panel : $dA_{\lambda}(t)/dt$ Vs. t. Note $dA_{\lambda}(t)/dt \rightarrow$ Const. $\Rightarrow$
normal diffusion.

\item Figure 7 : Upper Panel : Synthetic LVCF $R(\tau) \sim \textrm{e}^{-\tau/C_1} ~ \cos(\omega \tau)$
($C_1=7 , \omega=0.3 $).
Lower Panel : Induced AD.

\item Figure 8 : Upper Panel : Meridional AD (JJA and DJF curves have been shifted for clarity).
Ballistic $\rightarrow$ Subdiffusive $\rightarrow$ Diffusive.
Lower Panel : $A_z(t)/t \sim t \rightarrow A_z(t)/t \sim t^{\beta} ~-1<\beta<0 \rightarrow 
A_z(t)/t \sim$ Const.

\end{itemize}

\clearpage

\clearpage

\clearpage

\clearpage

\clearpage

\clearpage


\clearpage

\begin{thebibliography}{99}

\bibitem{Hinze} Hinze, J., 1975: {\em Turbulence}, McGraw-Hill, 790pp. 

\bibitem{Monin} Monin, A. and A. Yaglom, 1971: {\em Statistical Fluid Mechanics. Vol. 1}, 
MIT Press, 769pp. 

\bibitem{Shep_1} Shepherd, T., 1987: Rossby waves and two-dimensional turbulence in a 
large-scale zonal jet.
{\em Journal of Fluid Mech.}, {\bf 183}, 467-509.

\bibitem{Shep_2} Shepherd, T., 1987: A spectral view of nonlinear fluxes and stationary-
transient interaction in the atmosphere.
{\em J. Atmos. Sci.}, {\bf 44}, 1166-1179.

\bibitem{Taylor} Taylor, G., 1921: Diffusion by continuous movement.
{\em Proc. London Math. Soc.} , {\bf 20}, 196-211.

\bibitem{BG} Bouchaud, J-P. and A. Georges, 1990: Anomalous Diffusion in Disordered Media :
statistical Mechanisms, Models and Physical Applications.
{\em Physics Reports}, {\bf 195}, 127-293.

\bibitem{Zas} Leoncini, X. and G. Zaslavsky, 2002: Jets, stickiness, and anomalous transport.
{\em Physical Review E}, {\bf 65}, 046216. 

\bibitem{Boffetta} Boffetta, G., A. Celani, M. Cencini, G. Lacorata and A. Vulpiani, 2000: 
Nonasymptotic properties of transport and mixing.
{\em Chaos}, {\bf 10}, 50-60.

\bibitem{Giff1} Gifford, F., 1982: Horizontal diffusion in the atmosphere : A
Lagrangian-Dynamical theory.
{\em Atmospheric Environment}, {\bf 16}, 505-512.

\bibitem{Giff2} Gifford, F., 1984: The random force theory : Application to meso- and large-
scale atmospheric diffusion.
{\em Boundary Layer Meteorology}, {\bf 30}, 159-175.

\bibitem{Sawford} Sawford, B., 1984: The basis for, and some limitations of, the Langevin
equation in atmospheric relative dispersion modelling.
{\em Atmospheric Environment}, {\bf 18}, 2405-2411.

\bibitem{Pasquero} Pasquero, C., A. Provenzale and A. Babiano, 2001: Parameterization
of dispersion in two-dimensional turbulence.
{\em Journal of Fluid Mech.}, {\bf 439}, 279-303 

\bibitem{Pecseli} Pecseli, H. and J. Trulsen, 1997:  Eulerian and Lagrangian correlations in
two-dimensional random geostrophic flows.
{\em Journal of Fluid Mech.}, {\bf 338}, 249-276.

\bibitem{Mordant} Mordant, N., P. Metz, O. Michel and J.-F. Pinton, 2001: Measurement of Lagrangian Velocity in Fully
Developed Turbulence.
{\em Phys. Rev. Lett.}, {\bf 87}, 214501.

\bibitem{Pope} Pope, S., 2000: {\em Turbulent Flows}, Cambridge, 771pp. 

\bibitem{Berg} Berglund, A., 2004: Nonexponential statistics of flourescence photobleaching.
{\em Journal of Chemical Physics}, {\bf 121}, 2899-2903.

\bibitem{Anja} Feldmann, A. and W. Whitt, 1998: Fitting mixtures of exponentials to long-tail
distributions to analyze network performance models.
{\em Performance Evaluation}, {\bf 31}, 245-279.

\bibitem{Star} Starobinski, D. and M. Sidi, 2000: Modeling and analysis of power-tail
distributions via classical teletraffic methods.
{\em Queueing Systems}, {\bf 36}, 243-267.

\bibitem{Rhines} Rhines, P., 1994: Jets.
{\em Chaos}, {\bf 4}, 313-339.

\bibitem{Pedlosky} Pedlosky, J., 1987: {\em Geophysical Fluid Dynamics}
Springer Verlag, 710pp. 

\bibitem{Black} Blackmon, M., Y. Lee and J. Wallace, 1984: Horizontal Structure of the 
500 mb Height Fluctuations with Long, Intermediate and Short Time Scales.
{\em J. Atmos. Sci.}, {\bf 41}, 961-979.

\bibitem{Morel-D} Morel, P. and M. Desbois, 1974: Mean 200-mb Circulation in the 
Southern Hemisphere Deduced from EOLE Balloon Flights.
{\em J. Atmos. Sci.}, {\bf 31}, 394-407.

\bibitem{Weeks} Weeks, E., J. Urbach and H. Swinney, 1996: Anomalous diffusion on
asymmetric random walks with a quasi-geostrophic flow example.
{\em Physica D}, {\bf 97}, 291-310.

\bibitem{Flierl} Flierl , G., 1981: Particle Motions in Large-Amplitude Wave Fields.
{\em Geophys. Astrophys. Fluid Dynamics}, {\bf 18}, 39-74.

\bibitem{LaCasce} LaCasce, J.H. and K.G. Speer, 1999: Lagrangian statistics in unforced
barotropic flows.
{\em Journal of Marine Research}, {\bf 57}, 245-274.

\bibitem{Elliott}Elliott, F., D. Horntrop and A. Majda, 1997: Monte Carlo methods for turbulent tracers
with long range and fractal random velocity fields. {\em Chaos}, {\bf 7}, 39-48.

\bibitem{Basu} Basu, R., V. Naulin and J. Rasmussen, 2003: Particle diffusion in anisotropic turbulence.
{\em Communications in Nonlinear Science and Numerical Simulation}, {\bf 8}, 477-492.

\end{thebibliography}
\end{document}